\documentclass{article}
\usepackage[utf8]{inputenc}
\usepackage{amsmath,mathtools,amssymb}
\RequirePackage[numbers,sort&compress]{natbib}
\usepackage{graphics}
\usepackage{amsmath}
\usepackage{amssymb}
\usepackage{color}
\usepackage{graphicx}
\usepackage{caption,subcaption}
\usepackage{float}
\usepackage{authblk}

\title{Memory effects displayed in the evolution of continuous variable system}

\author{Samaneh Hesabi, Anindita Bera, and Dariusz Chru{\'s}ci{\'n}ski\\
Institute of Physics, Faculty of Physics, Astronomy and Informatics,
Nicolaus Copernicus University, Grudzi\c{a}dzka 5/7, 87--100 Toru{\'n}, Poland}

\begin{document}

\maketitle

\abstract{We analyze non-Markovian memory effects displayed by the quantum Brownian motion modelled as  quantum harmonic oscillators coupled to a bath consisting of harmonic oscillators. We study the time evolution of fidelity, Petz–R{\'e}nyi relative entropy and quantum entanglement for a family of 1-, 2- and 3-mode Gaussian states.}

\section{Introduction}

In recent years, open quantum systems \cite{bre} have received a lot of attention. This is due to the fact that any real quantum system is never perfectly isolated and the proper description of system's dynamics has to take into account the nontrivial interaction with its environment. Such interaction leads to well known phenomena like dissipation of energy and quantum decoherence \cite{zur,Sch,joos}. In the simplest scenario, when the interaction between the system and environment is  weak and their time scales are well separated, one applies  Markovian approximation giving rise to the celebrated Markovian master equation \cite{GKS,Lin}:

\begin{equation}\label{GKLS}
  \mathcal{L}(\rho) = -i[H_{S},\rho] + \sum_k \gamma_k \Big( L_k \rho L_k^\dagger - \frac 12 \{ L_k^\dagger L_k,\rho\} \Big) ,
\end{equation}
where $H_{S}$ stands for the effective system's Hamiltonian (including Lamb shift correction),  $L_k$ are jump (Lindblad) operators, and the transition rates $\gamma_k \geq 0$ (in what follows we keep $\hbar=1$). In non-Markovian regime,  the reduced evolution of the system is no longer governed by (\ref{GKLS}) (cf. recent reviews \cite{rivas,bre1,vega}). Usually, in the non-Markovian scenario one still uses time-local master equation where the system's Hamiltonian, Lindblad operators and transition rates are time dependent as well as the  rates $\gamma_k(t)$'s are temporally negative.  Such evolution displays characteristic memory effects leading for example to the well known phenomenon of information backflow \cite{bre1}. Note that if $H_S$, $L_k$ and $\gamma_k$ are just time dependent, and $\gamma_k(t)$'s are not negative,  the evolution is CP-divisible  and it is also considered as  the Markovian evolution \cite{rivas2}.

In this paper, we analyze non-Markovian evolution of one of paradigmatic models of open quantum systems \cite{bre}, namely, the quantum harmonic oscillators coupled to an ensemble of harmonic oscillator that is known as Quantum Brownian motion (QBM) \cite{Ein1,colla1,wei}. 
In particular, we analyze the time evolution of  fidelity \cite{banchiprl}, Petz-R{\'e}nyi relative entropy \cite{lami} and quantum entanglement \cite{Ade2} for Gaussian states. 
Both fidelity and relative entropy allow to to measure the distinguishability between quantum states. It is well known that for non-Markovian dynamics distingushability can temporally increase and this provide a clear sign of memory effects. The same applies for various correlation measures e.g. quantum entanglement.

In recent years, there has been special attention to the study of continuous variable systems especially Gaussian states, motivated by the fact that they are easily more accessible and controllable than discrete variable states \cite{Ade1,Chris,serabook}. An important subject is the study of the dynamics of quantum correlations of two mode Gaussian states in Markovian and non-Markovian open quantum systems \cite{mess3,sou,tor,hes1,hes2,hes3}. However, less attention has been paid to  three mode Gaussian states especially in the non-Markovian regime \cite{sab,vas1,dav,Alex,Isar,vas2}.

In this paper, we study  how the memory effects affect the  evolution of entanglement, fidelity  and  
R{\'e}nyi relative entropy of three classes of Gaussian states:  one and two mode squeezed and three mode basset-hound states as the initial states. For the single mode scenario, the state evolves with the QBM channel \cite{oli,wolf,Ade2}, whereas for more than one mode, only one mode interacts with QBM and the other modes do not transform.
Moreover, the effect of the environmental parameters on the evolution of entanglement, fidelity and Petz-R{\'e}nyi relative entropy is investigated.

The paper is structured as follows. Continuous variable systems are briefly reviewed in Sec.~\ref{cv1} where we also describe three classes of Gaussian states used in this paper. In Sec.~\ref{pm}, we introduce the physical model for the system and environment. 
The main results --- the evolution of entanglement, fidelity and Petz-R{\'e}nyi relative entropy --- are presented and discussed in detail in Sec.~\ref{res}. 
Finally, in Sec. ~\ref{conclude} we conclude.

\section{Preliminaries: CV systems}
\label{cv1}

A continuous variable (CV) system of $n$ canonical bosonic modes is described by a Hilbert space $\mathcal{H}= \bigotimes_{i=1}^n \mathcal{H}_i$ resulting from the tensor product structure of infinite-dimensional Hilbert spaces $\mathcal{H}_i$'s, each of them associated to a single mode \cite{plenio1,loock1,Ade1}. Let $a_i$ and $a_i^\dagger$ be the annihilation and creation operators acting on $ \mathcal{H}_i$, and $\hat{x}_i=(a_i+a_i^\dagger)$ and $ \hat{p}_i=(a_i-a_i^\dagger)/i$ be the related quadrature phase operators. Therefore the corresponding phase space variables are denoted by $x_i$ and $p_i$.
Let $\hat{X}=\left(\hat{x}_1,\hat{p}_1,\ldots,\hat{x}_n,\hat{p}_n\right)$ denotes the vector of the operators $\hat{x}_i$ and $\hat{p}_i$. 
The canonical commutation relations for the $\hat{X}_i$ can be expressed in terms of the symplectic form $\Omega$:
\begin{equation}
\left[\hat{X}_i,\hat{X}_j\right]=2 i \Omega _{\text{ij}},
\end{equation}
with
\begin{equation}
\Omega \equiv \bigoplus_{i=1}^n \omega,~~\omega \equiv \left(
\begin{array}{cc}
 0 & 1 \\
 -1 & 0 \\
\end{array}
\right),
\end{equation}
where $\omega$ is usually referred to as the one-mode standard symplectic form.
The state of a CVs system can be equivalently described by a positive trace-class operator (the density matrix) or by quasi-probability distributions.
States with Gaussian characteristic functions and quasi-probability distributions are referred to as Gaussian states. Such states are at the heart of information processing in CVs
systems and are the main interest of our paper. From the definition, a Gaussian state is completely characterized by the first and second statistical moments of the quadrature field operators.  The vector of first moments is denoted by  $\bar X \equiv (\langle {\hat X}_1 \rangle,\langle {\hat X}_2 \rangle,\ldots,\langle {\hat X}_{2n-1} \rangle,\langle {\hat X}_{2n} \rangle)$ and second moment is a $2n \times 2n$ matrix, namely, covariance matrix (CM) $\sigma$  of elements \cite{wolf,Ade2,serabook,Ade3}:
\begin{equation}
\sigma _{{\rm{ij}}} = \frac{1}{2}\langle {\hat X_i}{\hat X_j} + {\hat X_j}{\hat X_i}\rangle  - \langle {\hat X_i}\rangle \langle {\hat X_j}\rangle.
\end{equation}
Note that without loss of generality, one can consider the first moments as null,
and then, for all the informational purposes, any Gaussian state can
be completely determined by its covariance matrix.
Now $\sigma$ will be a bonafide CM if it fulfills the Robertson-Schr\"odinger uncertainty relation \cite{simon1,simon2,wolf,Ade2,serabook}:

\begin{equation}
\label{ur1}
\sigma +i \Omega \geq 0.
\end{equation}
For any physical CM $\sigma$, there exist a symplectic transformation $S \in Sp_{2n,\mathbb{R}}$\footnote{Symplectic transformations on a $2n$ dimensional phase space form the (real) symplectic group is denoted by $Sp_{2n,\mathbb{R}}$. Such transformations act on a CM  by congruence:
$\sigma \rightarrow S^T \sigma S$. Also $\mbox{Det}~ S = 1~ \forall~ S \in Sp_{2n,\mathbb{R}}.$}
such that $S^T \sigma S = \nu$ where
\begin{equation}
    \nu =\underset{k=1}{\overset{n}{\oplus }}\left(
\begin{array}{cc}
 \nu _k & 0 \\
 0 & \nu _k \\
\end{array}
\right).
\end{equation}
The quantities $\{\nu _k\}$ constitute the symplectic spectrum of $\sigma$ and $\nu$ is said to be the  Williamson normal form associated with $\sigma$ \cite{john,simon3}. Note that the fundamental properties such as the uncertainty relation for Gaussian states can be easily expressed in terms of the symplectic eigenvalues $\nu_k$. Therefore  from Eq.~(\ref{ur1}) one can easily get:
\begin{equation}
\label{ev1}
     \nu _k\geq 1.
\end{equation}

For our convenience, let us write the CM $\sigma_{1 \ldots n}$ of an n-mode Gaussian state which can be expressed in terms of $2 \times 2$ sub-matrices of each mode in the following way \cite{Ade4}
\begin{equation}
\label{nmode}
    \sigma_{1 \ldots n} = 
 \begin{pmatrix}
  \sigma_{1} & \varepsilon_{12} & \cdots & \varepsilon_{1n} \\
  \varepsilon_{12}^T & \ddots & \ddots & \vdots \\
  \vdots  & \ddots  & \ddots & \varepsilon_{n-1,n}  \\
  \varepsilon_{1n}^T &\cdots & \varepsilon_{n-1,n}^T & \sigma_{n} 
 \end{pmatrix}.
\end{equation}
Each diagonal block $\sigma_k$ is respectively the local CM corresponding to the reduced state of mode $k$, for all $k=1,\dots,n$.  On the other hand, the off-diagonal matrices $\varepsilon_{i,j}$ encode the intermodal correlations (quantum and classical) between subsystems $i$ and $j$. For product states, all the matrices $\varepsilon_{i,j}$ vanish.

Let us briefly introduce three classes of Gaussian states which are important in our analysis: one, two and three mode Gaussian states, as the initial states of the system. 
In particular, from one mode Gaussian states, we consider the single mode squeezed states 
$|r\rangle =\hat{U}(r)|0\rangle$  with squeezing factor $r \in \mathbb{R}$, where  $\hat{U}(r)=\exp \big[-\frac{r}{2} (\hat{a}_k^{\dagger^2} -\hat{a}_k^2  ) \big]$.
The covariance matrix of single mode squeezed state is given by
\begin{equation}
 \label{sq1}
        \sigma_1^{sq} = \left( \begin{array}{cc}
        e^{-2r} & 0\\
        0 & e^{2r} 
    \end{array} \right).
\end{equation}

Following the Eq.~(\ref{nmode}), the form of CM of a two-mode Gaussian states is as follows:
\begin{equation}
\sigma_{12} =\left(
\begin{array}{ccc}
 \sigma _1 & \varepsilon _{12}\\
 \varepsilon _{12}^T & \sigma _2 \\
\end{array}
\right),
\label{2modecm}
\end{equation}

where $\sigma _1$ and $\sigma _2$ are the covariance matrices corresponding to each mode and $\varepsilon _{12}$ is the correlation matrix between them. In this paper, we consider
an important instance of two mode Gaussian state, namely, the two-mode squeezed states $|\psi^{sq}\rangle_{ij}=\hat{U}_{ij}(r) (|0\rangle_i \otimes |0\rangle_j)$ with squeezing factor $r \in \mathbb{R}$,  where the (phase-free) two-mode squeezing operator is given by
\begin{equation}
    \hat{U}_{ij}(r)=\exp \big[-\frac{r}{2} (\hat{a}_i^\dagger \hat{a}_j^\dagger-\hat{a}_i \hat{a}_j )\big].
\end{equation}
It is important to mention here that these states are the key resources for practical implementations of CV quantum information protocols \cite{loock1}. A two-mode squeezed
state with squeezing parameter $r$, also known in quantum optics as a twin-beam state, takes  the following form of CM
 \begin{equation}
 \label{sq}
        \sigma_{ij}^{sq} = \left( \begin{array}{cc|cc}
        \cosh{2r} & 0 & \sinh{2r} & 0\\[1ex]
        0 & \cosh{2r} & 0 & -\sinh{2r}\\[1ex]
        \hline
        \sinh{2r} & 0 & \cosh{2r} & 0\\[1ex]
        0 & -\sinh{2r} & 0 & \cosh{2r}\\[1ex]
    \end{array} \right)\!.
    \end{equation}

Three mode Gaussian states \cite{Ade4} can be classified into two classes of states endowed with
symmetries under mode exchange:  fully symmetric class of states  which are invariant under the permutation of all modes and  bisymmetric class of states that are invariant under the exchange of a specific pair of modes. It is  obvious that fully symmetric states are bisymmetric under any bipartition of the modes. 
In this study, we consider the bisymmetric Gaussian states, which is also known as the basset hound states \cite{sera1,g1}, described by the following CV:
\begin{equation}
\label{bh1}
\sigma_{bh} =\left(
\begin{array}{ccc}
 \sigma _1 & \varepsilon _{12} & \varepsilon _{13} \\
 \varepsilon _{12}^T & \sigma _2 & \varepsilon _{23} \\
 \varepsilon _{13}^T & \varepsilon _{23}^T & \sigma _3 \\
\end{array}
\right),
\end{equation}
where 

$$ \sigma _1=a \mathbb{I}_2 , \ \ \sigma _2=\sigma _3=\frac{(a+1)}{2} \mathbb{I}_2 , \ \ \varepsilon _{23}=\frac{(a-1)}{2}\mathbb{I}_2 , $$
and
$$ \varepsilon _{12}=\varepsilon_{13}=\frac{\sqrt{a^2-1}}{\sqrt{2}} \, \left(
\begin{array}{cc} 1 & 0 \\ 0 & -1 \end{array} \right)  , $$
with $a=\cosh{2r}$.


\section{Physical model}
\label{pm}

In this section we describe the physical model that we wish to study.
We consider Quantum Brownian motion (QBM) which describes a linear interaction of a quantum harmonic oscillator with unit mass and frequency $\omega _0$ as system with a bosonic environment.
The bosonic environment is a bath consisting an ensemble of harmonic oscillators with
masses $m_n$ and frequencies $w_n$.
The Hamiltonian of the total system of oscillators is given by \cite{bre}:
\begin{equation}
\hat{H}=\frac{\hat{p}^2}{2}+\frac{1}{2} \hat{q}^2 \omega _0^2+\sum _n \left(\frac{\hat{P}_n^2}{2 m_n}+\frac{1}{2} m_n \omega _n^2 \hat{Q}_n^2 \right)+\alpha \hat{q} \sum _n K_n \hat{Q}_n,
\end{equation}
where $\hat{q} (Q_n)$ and $\hat{p} (P_n)$ are the position and momentum of the system (environment). $K_n$ and $\alpha$ are the relative strengths of the interaction and the coupling constant, respectively. With the assumption of weak coupling
between system and environment and secular approximation, the master equation for this model in the interaction picture is given by \cite{mess1,mess2}:
\begin{equation}
\label{mastereq}
\frac{d \rho }{d t}=\frac{\Delta(t)+\gamma(t)}{2} (2 \hat{a} \rho \hat{a}^{\dagger}-\rho \hat{a}^{\dagger }\hat{a}-\hat{a}^{\dagger}\hat{a} \rho)+\frac{\Delta(t)-\gamma (t)}{2} (2 \hat{a}^{\dagger }\rho  \hat{a}  -\hat{a} \hat{a}^{\dagger } \rho-\rho \hat{a} \hat{ a}^{\dagger }),
\end{equation}
where the time-dependent terms $\Delta(t)$ and $\gamma(t)$ are the diffusion and damping term, respectively. In the case of a thermal environment at temperature T, the coefficients of $\Delta(t)$ and $\gamma(t)$ are in the following form \cite{mess2}:
\begin{equation}
\Delta (t)=\alpha ^2\int _0^t d\tau  \int _0^{\infty } d\omega  J (\omega)    \coth (\frac{\hbar \omega }{2k T})\cos (\omega \tau) \cos (\omega _0\tau),
\end{equation}
and
\begin{equation}
\gamma (t)=\alpha ^2\int _0^t d\tau\int _0^{\infty }   d\omega  J(\omega)  \sin(\omega \tau )   \sin  (\omega _0 \tau).
\end{equation}
Here $ J (\omega)$ characterizes the spectral density of the environment. In particular, in our study, we consider an Ohmic spectral density with Lorentz-Drude cutoff \cite{bre,mess3}:
\begin{equation}
\label{ohmic}
J (\omega )=\frac {2 \omega}{\pi } \frac{\omega _c^2} {\omega _c^2+\omega _0^2},
\end{equation}
where $\omega_c$ is the cutoff frequency. Therefore the expressions for $\gamma(t)$ and $\Delta(t)$ read \cite{mess3}:
\begin{equation}
    \gamma (t)=\frac{\alpha ^2 x^2 \omega _0}{x^2+1} \Big(1-e^{-\omega _c t}\cos( \omega _0 t)-x e^{-\omega _c t}\sin ( \omega _0 t) \Big),
\end{equation}
and 
\begin{equation}
\begin{array}{*{20}{l}}
\begin{array}{l}
\Delta (t) = \frac{{\alpha ^2}{x^2}{\omega _0}}{x^2 + 1}\Big(\coth(\pi {r_0})- \cot(\pi {r_c}){e^{-\omega_c t}}[r\cos ({\omega _0}t)-\sin ({\omega _0}t)]\\
\end{array}\\
\begin{array}{l}\\\,\,\,\,\,\,\,\,\,\,\,\,+ \frac{1}{\pi {r_0}}\cos ({\omega _0}t)[\bar F(-{r_c},t) + \bar F({r_c},t) - \bar F(i{r_0},t) - \bar F(-i{r_0},t]\\
\end{array}\\
\begin{array}{l}\\\,\,\,\,\,\,\,\,\,\,\,\,-\frac{1}{\pi}
\sin ({\omega _0}t)\Big[\frac{e^{-\nu _1t}}{2r_0(r_0^2+1)}[(r_0-i) \bar G(-r_0,t)+(r_0+i)\bar G(r_0,t)] \\
\end{array}\\
\\\,\,\,\,\,\,\,\,\,\,\,\,
+\frac{1}{2r_c}[\bar F(-r_c,t)-\bar F(r_c,t)]\Big]\Big).
\end{array} 
\end{equation}
In the above equation, we take $x = \omega _c /\omega _0$, $r_0=\omega_0/2 \pi T$, $r_c=\omega_c/2 \pi T$, $\nu _1=2\pi k T$, and
\begin{equation}
 \bar{F}(x,t)= {}_2F_1(x,1,x+1,e^{-\nu _1 t}),  
\end{equation}
\begin{equation}
\bar{G}(x,t)={}_2F_1(2,x+1,x+2,e^{-\nu _1 t}),
\end{equation}
where ${}_2F_1(a,b,c,z)$ is the hypergeometric function \cite{gra}.

In our analysis, we study the evolution of an multimode Gaussian state in the following way. When the initial single mode Gaussian state $\sigma(0)$ is subjected to the QBM channel described in Eq.~(\ref{mastereq}), then the evolved covariance matrix takes the following form:
\begin{equation}
\label{evo2}
    \sigma(t)= 
 e^{-\Gamma(t)} \sigma(0)+2\Delta_\Gamma(t)\mathbb{I},
\end{equation}
where $\Gamma(t)=\int_0^t2\gamma(s)ds$ and $\Delta_{\Gamma}(t)=e^{-\Gamma(t)}\int_0^{t}{e^{\Gamma(s)}\Delta(s)}ds$.
For more than one mode, we consider an initial $n$-mode covariance matrix $\sigma _{n}(0)$, and then the first mode undergoes a QBM evolution under the master equation given by Eq.~(\ref{mastereq}) and other modes are subjected to the free unitary evolutions. Hence the evolved CM can be written as \cite{mess3}:
\begin{eqnarray}
\label{evo1}
\sigma _{n}(t) = {[{e^{ -\frac{{\Gamma (t)}}{2}}}{\mathbb{I}{_A}} \oplus {\mathbb{I}}{_B} \oplus {\mathbb{I}{_C}} \oplus \ldots ]^T}{\sigma _{n}}(0)[{e^{ - \frac{{\Gamma (t)}}{2}}}{\mathbb{I}{_A}} \oplus {\mathbb{I}{_B}} \oplus {\mathbb{I}{_C}} \oplus \ldots ]  \nonumber\\
 +\Delta _{\Gamma }(t)[{\mathbb{I}{_A}} \oplus {\mathbb{O}{_B}} \oplus {\mathbb{O}{_C}} \oplus \ldots].
\end{eqnarray}
Note that in the weak coupling regime $\alpha \ll 1 $ that was already incorporated
in the derivation of the master equation, one can expand  $\Delta_{\Gamma}$ to first order in $\Gamma(t)$. Since  $\Gamma(t) \propto \alpha^2$ and $\Delta(t) \propto \alpha^2$ in the weak coupling regime, then by truncating the expansion up to second order in $\alpha$, one can  write $\Delta_{\Gamma}(t) \simeq \int_0^t \Delta(s) ds$ \cite{vas1}.

\begin{figure}[!ht]
    \centering
    \includegraphics[scale=0.7]{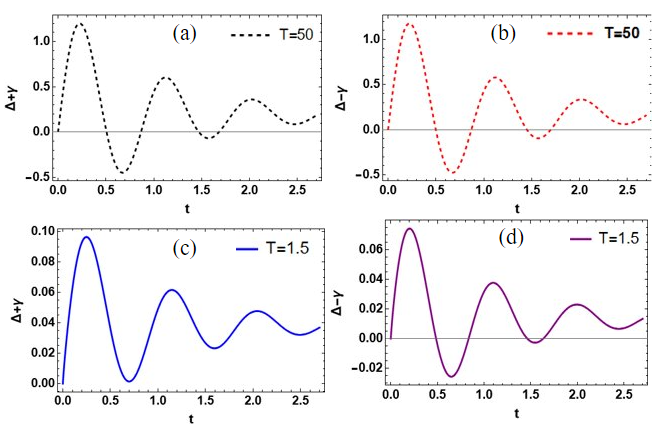}
    \caption{The time evolution of $\Delta+\gamma$ and $\Delta-\gamma$ with low and high temperatures. In the plots (a) and (b), we consider moderately high temperature $T=50,$  whereas plots (c) and (d) consist of low temperature value $T=1.5$. Here, $\alpha=0.3$, $\omega_0=7$, and $\omega_c=1$.}
    \label{fig:1}
\end{figure}

In Fig. \ref{fig:1}, we show the time evolution of coefficients $\Delta$+$\gamma$ and $\Delta$-$\gamma$ for both high and low temperatures. 
As we observes, in the initial times, i.e. for $t \omega_c < 2$, these coefficients $\Delta \pm \gamma$ can acquire temporarily negative values.
It proves that the evolution is non-Markovian and one can expect typical memory effects. Note that at high temperature, $\Delta(t) \gg \gamma(t)$, and therefore the time evolution of $\Delta \pm \gamma$ looks almost similar. Moreover, for the specific Ohmic spectral density that we considered in Eq. (\ref{ohmic}), the non-Markovianity in the QBM channel depends on the ratio between the characteristic frequency of the system $\omega_0$ and the cutoff frequency $\omega_c$, i.e.,
$\omega_0/\omega_c$. When $\omega_0/\omega_c \ll 1$, the
dynamics of the system is essentially Markovian for any temperature $T$ \cite{mess3,sou,tor}. But in the regime $\omega_0/\omega_c > 1$, one can see the
non monotonic behavior implying the evolution is non-Markovian. However, asymptotically the  evolution, irrespective of the system and bath parameters, all memory effects  vanish \cite{mess3,sou,tor}. In our work, we investigate the non-Markovian behavior in QBM for both the low and  high temperature regime
 with $\omega_0/\omega_c > 1$.

\section{Memory effects }
\label{res}
To study the  non-Markovian behavior in QBM, we investigate the time evolution of entanglement, fidelity
and  relative entropy.

\subsection{Evolution of fidelity}
In Quantum information theory, fidelity provides a measure of the distinguishability between two quantum states.  The fidelity of two quantum states $\rho _1$ and $\rho _2$ is defined as \cite{Armin}:
\begin{equation}
F\left(\rho _1,\rho _2\right)=\left[\text{Tr}\left(\sqrt{\sqrt{\rho _1} \rho _2 \sqrt{\rho _1}}\right)\right]{}^2.
\end{equation}

The authors in Ref.~\cite{banchiprl} showed that the quantum fidelity between two arbitrary multimode Gaussian states can be expressed in terms of the relative average displacement 
$\delta \left\langle\hat{\mu}\right\rangle:=\left\langle\hat{\mu}\right\rangle_{\rho _1}-\left\langle\hat{\mu}\right\rangle_{\rho _2}$ and their CMs $V_1$ and $V_2$ in the following way:
\begin{equation}
\label{fidgen}
    F\left(\rho _1,\rho _2\right) = \Big[F_0(V_1,V_2)\exp \big[-\frac{1}{4}(\delta \langle \hat{\mu}\rangle )^T (V_1+V_2)^{-1} \delta\langle \hat{\mu }\rangle \big] \Big]^2. 
\end{equation}
Importantly, here the symplectic eigenvalues of CMs $V_1$ and $V_2$ are greater or equal to $1/2$. The authors in Ref.~\cite{banchiprl} also considered a modified version of the CM: $ \sigma =-2Vi \Omega$ such that the symplectic eigenvalues of $\sigma$ are greater or equal to 1, satisfying Eq.~(\ref{ev1}). 
 For two arbitrary Gaussian states, the term $F_0(V_1,V_2)$ can be written as \cite{banchiprl}:
 \begin{equation}
   F_0(V_1,V_2)=\frac{F_{tot}}{\sqrt[4]{\det(V_1+V_2)}},
\end{equation}
where
\begin{eqnarray}
    F_{tot}^4 &=& \det \left[2\big(\sqrt{\mathbb{I}+\frac{(V_{aux}\Omega)^{-2}}{4}}+\mathbb{I}\big) V_{aux}\right] \\
    &=& \det \left[\big(\sqrt{\mathbb{I}-\sigma_{aux}^{2}}+\mathbb{I}\big)\sigma_{aux}i\Omega\right].
\end{eqnarray}
The auxiliary matrices can be written as: 
\begin{equation}
   V_{aux} = \Omega^T(V_1+V_2)^{-1}(\frac{\Omega}{4}+V_2 \Omega V_1),  
\end{equation}
and 
\begin{equation}
   \sigma_{aux} = -2V_{aux}i\Omega.  
\end{equation}

Note that in this work, we consider undisplaced modes, therefore the exponential
term in Eq.~(\ref{fidgen}) becomes 1. The fidelity for single and two mode has simplified form \cite{mari} and it can be expressed in terms of their CMs $\sigma_1$ and $ \sigma_2$.
For one mode Gaussian states: 
\begin{equation}
    F(\rho_1, \rho_2)=\frac{1}{\sqrt{\Sigma+\Lambda}-\sqrt{\Lambda}},
\end{equation}
and for two mode Gaussian states:
\begin{equation}
\label{fid1}
 F(\rho_1, \rho_2)=\frac{1}{\sqrt{\eta }+\sqrt{\Lambda }-\sqrt{(\sqrt{\eta }+\sqrt{\Lambda })^2-\Sigma}},
\end{equation}
where 
\begin{equation}
\Sigma =\det \left(\sigma _1+\sigma _2\right)\geq 1,
\end{equation}
\begin{equation}
\eta =2^{2 n} \det \left[\left(\Omega \sigma _1  \right) \left(\Omega \sigma _2  \right)-\frac{\mathbb{I}}{4}\right]\geq \Sigma,
\end{equation}
\begin{equation}
\Lambda = 2^{2 n} \det\left(\sigma _1+\frac{i \Omega }{2}\right)  \det\left(\sigma _2+\frac{i \Omega }{2}\right)\geq 0.
\end{equation}

\begin{figure}[!ht]
    \centering
    \includegraphics[scale=0.6]{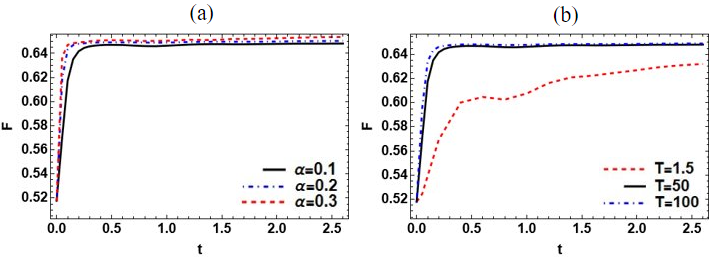}
    \caption{The evolution of fidelity versus time for one-mode squeezed states with (a) different coupling constant for $T=50$, and (b) different temperature for $\alpha=0.1$. Here $\omega_o=7$, $\omega_c=1$, $r_1=2$, $r_2=3$.}
    \label{fig:2}
\end{figure}

 At first, we assume the one mode squeezed states as the initial states with different squeezing parameters $r_1$ and $r_2$ and then we evaluate the fidelity with time for different values of coupling constant $\alpha$ and temperature $T$, which has been shown in Figs. \ref{fig:2}(a) and \ref{fig:2}(b). 
 Fig.  \ref{fig:2} shows that fidelity is not monotonic for $t\omega_c \in (0,1.8)$. Clearly, this behaviour is compatible with negativity of $\Delta \pm \gamma$   (see Fig.~\ref{fig:1}). However, with the increasing $t$, fidelity is an increasing function of both $\alpha$ and $T$,  and eventually it saturates to a constant value. Moreover, from  Fig. \ref{fig:2}(b), we  observe that with increasing temperature, the fidelity becomes almost constant at short time.

\begin{figure}[!ht]
    \centering
    \includegraphics[scale=0.5]{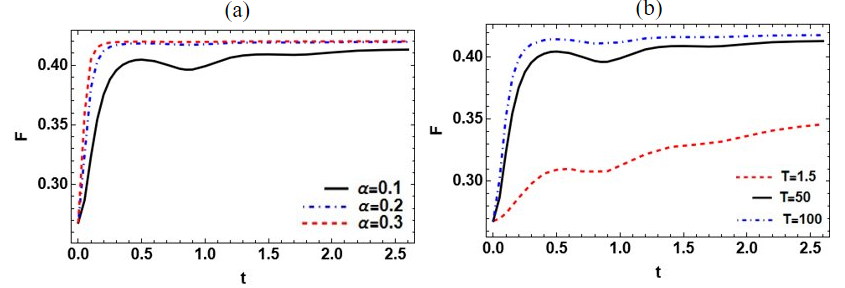}
    \caption{The evolution of fidelity two-mode squeezed states versus time with (a) different coupling constant for $T=50$, and (b) different temperature  for $\alpha=0.1$. Here 
     $\omega_o=7$, $\omega_c=1$, $r_1=2$, $r_2=3$.}
    \label{fig:3}
\end{figure}
 
In Fig. \ref{fig:3}, we perform the similar analysis as above but with the two-mode squeezed and three mode basset-hound states as the initial states with different squeezing parameters $r_1$ and $r_2$. However, we find that the behavior of fidelity for both the states are essentially the same. Therefore we plot fidelity with time in Fig.  \ref{fig:3} only for two-mode squeezed states. 
 We find that in the initial time, the value of fidelity is not monotonic implying the fact that the evolution is non-Markovian. However, with the time, fidelity increases with increasing $\alpha$ and $T$ and eventually it saturates.
 Note that compare to the single-mode squeezed states, the non-Markovian behavior of fidelity  with time enhances for two mode squeezed states.

\subsection{Evolution of entanglement}

We use negativity as a suitable measure of quantum entanglement. Negativity for a Gaussian state with CM $\sigma$ is given by \cite{Ade2}:

\begin{equation}
   \mathcal{N}(\sigma)=
  \begin{cases}
    \frac{1}{2} (\Pi_k  \tilde{\nu}_k^{-1} -1)      & \quad \text{for } k: \tilde{\nu}_k <1,\\
    0 & \quad \text{if}~\tilde{\nu}_i>1~ \forall i.
  \end{cases}
\end{equation}
Here $\{\tilde{\nu}_k\}$ represents the set of symplectic eigenvalues of the partially transposed 
CM $\tilde{\sigma}$. Accordingly, the logarithmic negativity is:
\begin{equation}
\label{ln}
    E_{\mathcal{N}}(\sigma)=
  \begin{cases}
    -\sum _{k} \log  \tilde{\nu }_k       & \quad \text{for } k: \tilde{\nu}_k <1,\\
    0 & \quad \text{if}~\tilde{\nu}_i>1~ \forall i.
  \end{cases}
\end{equation}

It turns out that for two mode CM in Eq.~ (\ref{2modecm}), one can write the symplectic eigenvalues of the partial transpose CM $\tilde \sigma$  \cite{ade}:
\begin{equation}
{\tilde \nu }_ \pm = \sqrt {\frac{{\tilde \Delta  \pm \sqrt {{{\tilde \Delta }^2} - 4\det [\sigma ]} }}{2}},
\end{equation}
where $\tilde \Delta=\det[\sigma_1]+\det[\sigma_2]-2 \det[\varepsilon_{12}]$.

We evaluate the entanglement i.e. logarithmic negativity, given in Eq.~(\ref{ln}) for a two-mode squeezed and three mode basset-hound state. Let us take $r=2$.
 \begin{figure}[h]
    \centering
    \includegraphics[scale=0.47]{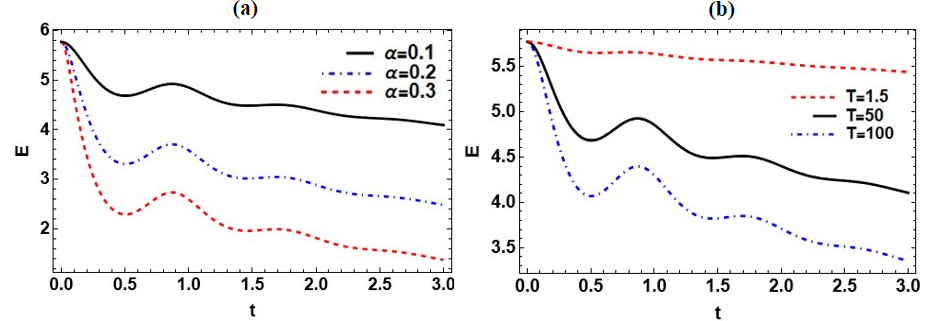}
    \caption{The evolution of entanglement of two-mode squeezed states versus time with
    (a) different coupling constant for $T=50$, and (b) different temperature for $\alpha=0.1$. Here $\omega_o=7$, $\omega_c=1, r=2$.}
    \label{fig:4}
\end{figure}

In Fig. \ref{fig:4}, we plot the evolution of entanglement with time for two-mode squeezed states for different values of coupling constant $\alpha$ and temperature $T$. From both the Figs. \ref{fig:4}(a) and \ref{fig:4}(b), we can see that the value of entanglement decreases with increasing $\alpha$ and $T$. The evolution of entanglement with time for three mode basset-hound states is essentially same with that for the two-mode squeezed states. Clearly from  the Fig. \ref{fig:4}, it is evident that in the initial time interval, one can observe the non monotonicity behavior of entanglement which implies the evolution is non-Markovian.

\subsection{Evolution of Petz–R{\'e}nyi relative entropy}

For two Gaussian states $\rho$ and $\rho'$, the Petz–R{\'e}nyi relative entropy with parameter $\kappa \in (0,1) \cup (1,\infty )$ is defined as \cite{petz, lami}:
\begin{equation}
   D_\kappa (\left. \rho  \right\|\rho ') = \frac{1}{{\kappa  - 1}}\,\ln \,{Q_\kappa }(\left. \rho  \right\|\rho '),
\end{equation}
where ${Q_\kappa }(\left. \rho  \right\|\rho ')$ is the Petz–R{\'e}nyi relative quasi-entropy expressed via
\begin{equation}
Q_\kappa (\left. \rho  \right\|\rho ') \equiv \mbox{Tr} \big[ {\rho ^\kappa }{{\rho '}^{1 - \kappa }} \big].
\end{equation}
Here we consider $\kappa  \in (1,\infty )$. Let $\sigma_{\rho}$ and $\sigma_{\rho'}$ be the CMs of the Gaussian states $\rho$ and $\rho'$, respectively  such that
the following condition has to be satisfied:
\begin{equation}
\label{conentropy}
    \sigma _{\rho'(\kappa - 1)} > {\sigma _{\rho (\kappa )}},
\end{equation}
 where
\begin{equation}
 \sigma _{\rho (\kappa )} = \frac{{{{(\mathbb{I} + {{({\sigma _\rho }\,i\,\Omega )}^{ - 1}})}^\kappa } + {{(\mathbb{I} - {{({\sigma _\rho }\,i\,\Omega )}^{ - 1}})}^\kappa }}}{{{{(\mathbb{I} + {{({\sigma _\rho }\,i\,\Omega )}^{ - 1}})}^\kappa } - {{(\mathbb{I} - {{({\sigma _\rho }\,i\,\Omega )}^{ - 1}})}^\kappa }}}\,i\Omega,
\end{equation}
\begin{equation}
\sigma _{\rho '(\kappa  - 1)} = \frac{{{{(\mathbb{I} + {{({\sigma _{\rho '}}\,i\,\Omega )}^{ - 1}})}^{\kappa  - 1}} + {{(\mathbb{I} - {{({\sigma _{\rho '}}\,i\,\Omega )}^{ - 1}})}^{\kappa  - 1}}}}{{{{(\mathbb{I} + {{({\sigma _{\rho '}}\,i\,\Omega )}^{ - 1}})}^{\kappa  - 1}} - {{(\mathbb{I} - {{({\sigma _{\rho '}}\,i\,\Omega )}^{ - 1}})}^{\kappa  - 1}}}}\,i\Omega .
\end{equation}
Then  the Petz–R{\'e}nyi relative entropy is defined as follows \cite{lami}:

\begin{eqnarray}
\label{entropy1}
Q_\kappa (\left. \rho  \right\|\rho')= 
 \frac{{Z_{\rho'}^{\kappa  - 1}}}{{Z_{\rho}^\kappa }}\frac{{{Z_{\rho (\kappa )}}{Z_{\rho '(\kappa  - 1)}}}}{{{{[\det ([{\sigma _{\rho '(\kappa  - 1)}} - {\sigma _{\rho (\kappa )}}]/2)]}^{1/2}}}} \times \nonumber\\
 \exp \big[\delta \mu^T{({\sigma _{\rho '(\kappa  - 1)}} - {\sigma _{\rho (\kappa )}})^{ - 1}} \delta \mu \big],
\end{eqnarray}
where

\begin{equation}
Z_{\rho (\kappa )}= \sqrt {\det ([{\sigma _{\rho (\kappa )}}+i\Omega ]/2)},
\end{equation}
and

\begin{equation}
Z_{\rho '(\kappa  - 1)}=\sqrt {\det ([{\sigma _{\rho '(\kappa  - 1)}}+i\Omega ]/2)},
\end{equation}
\begin{equation}
 \delta \mu=\langle \mu\rangle_\rho-\langle \mu \rangle_{\rho'}.
\end{equation}

\begin{figure}[t]
    \centering
    \includegraphics[scale=0.6]{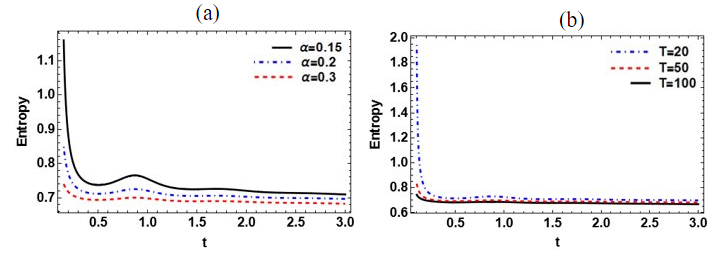}
    \caption{The evolution of the Petz–R{\'e}nyi relative entropy of one-mode squeezed states versus time for (a) different coupling constant for $T=50$, and (b) different temperature for $\alpha=0.3$. Here $\kappa=2$, $\omega_0=7$, $\omega_c=1$, $r_1=2$ and $r_2=3$.}
    \label{fig:6}
\end{figure}
Note that in this work, we consider undisplaced modes, therefore the exponential
term in Eq.~(\ref{entropy1}) becomes 1. Also, to calculate the entropy,  the condition in Eq.~(\ref{conentropy}) must be satisfied by the CMs of two Gaussian states $\rho$ and $\rho'$. It provides a critical value $t_*$ for time, i.e. only for $t > t_*$ the condition (\ref{conentropy}) holds. Clearly, $t_*$ depends upon $T$ and $\alpha$. We find that with the increasing values of $T$ and $\alpha$, this critical time $t_*$ decreases.  For instance, for one mode squeezes states if we take $T=50$ and $100$, then for $\alpha=0.15$, this condition satisfies when $t_*=0.13$ and $0.09$, respectively. However, for $\alpha=0.3$, one finds $t_*= 0.06$ and $0.05$, respectively. 
In Fig. \ref{fig:6}, we plot the evolution the Petz–R{\'e}nyi relative entropy with time for one-mode squeezed state for different values of coupling constant $\alpha$ and temperature $T$.
Note that to plot the entropy, we consider the maximum time scale satisfying condition (\ref{conentropy}) for different $\alpha$ and $T$.
As we observe,  the value of entropy decreases with increasing time. However, in the initial time, while decreasing, the entropy shows non monotonic behavior implying that the evolution is non-Markovian. 

\begin{figure}[h]
    \centering
    \includegraphics[scale=0.55]{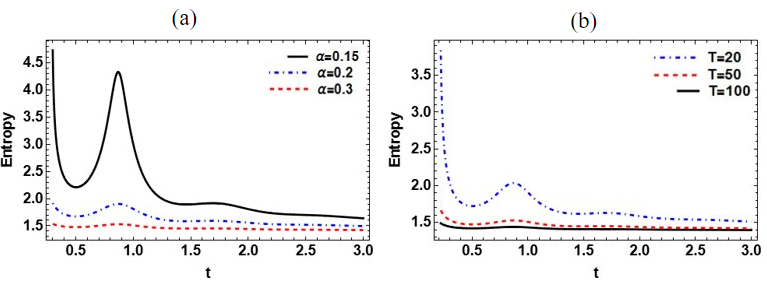}
    \caption{The evolution of the Petz–R{\'e}nyi relative entropy of two-mode squeezed states versus time for (a) different coupling constant for $T=50$, and (b) different temperature for $\alpha=0.3$. Here $\kappa=2$, $\omega_o=7$, $\omega_c=1$, $r_1=2$ and $r_2=3$.}
    \label{fig:7}
\end{figure}

Similarly, for the two mode squeezes states, to satisfy condition (\ref{conentropy}), 
at $T=50$ and $100$, we need to consider  $t_*= 0.3$ and $0.19$, respectively for $\alpha=0.15$. 
However, if we increase $\alpha$, say, $0.3$, then $t$ must be greater or equal to $t_*=0.13$ and $0.09$, respecively. In Fig. \ref{fig:7}, we plot the evolution the Petz–R{\'e}nyi relative entropy with time for two-mode squeezed state for different values of coupling constant $\alpha$ and temperature $T$.
Here also like the single mode, we consider the maximum time scale satisfying condition (\ref{conentropy}) for different $\alpha$ and $T$.
Clearly we can see that the entropy is a decreasing function of coupling constants and temperature.
However, from one mode to two mode squeezed states, the non monotonic behavior or entropy increases implying the fact that the evolution is non-Markovian. 
\begin{figure}[!ht]
    \centering
    \includegraphics[scale=0.65]{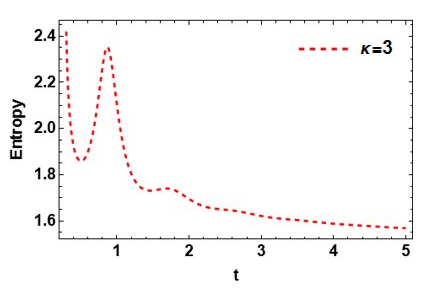}
    \caption{The evolution of the Petz–R{\'e}nyi relative entropy of two-mode squeezed state versus time for $\kappa=3$. Here $T=50$, $\alpha=0.3$, $\omega_0=7$, $\omega_c=1$, $r_1=2$ and $r_2=3$.}
    \label{fig:9}
\end{figure}

We also observe the behavior of  Petz–R{\'e}nyi relative entropy for a different value of $\kappa$ for two mode squeezed states  in Fig. \ref{fig:9}.
Clearly we see that in the initial time for $\kappa=3$, the entropy shows non monotonic behavior which implies the non-Markovianity of the evolution. However, with increasing $t$, entropy is a decreasing function.

 \section{Conclusions}
 \label{conclude}
 
In this work,  we addressed the time evolution of fidelity, relative entropy and quantum entanglement  in continuous variable system scenario. We considered one and two-mode squeezed  and three mode basset-hound state as the initial states and the quantum Brownian motion as a Gaussian map. The numerical results coherently witness the non-Markovian memory effects displayed in the initial time window where the parameters of the corresponding master equations are temporally negative. 
Outside this region the dynamics of the system leads to perfectly monotonic evolution of fidelity, relative entropy and entanglement.

\section*{Acknowledgements}

This paper was partially supported by the National Science Center project No 2018/30/A/ST2/00837. A. B. acknowledges conversation with Saptarshi Roy. We thank Sagnik Chakraborty for reading the manuscript and providing useful suggestions.

 \bibliographystyle{apsrev4-1}
\bibliography{name}
\end{document}